# Interface-Controlled Antiferromagnetic Tunnel Junctions based on a metallic van der Waals A-type Antiferromagnet

Wei-Min Zhao[1, †], Yi-Lun Liu[1, †], Liu Yang[2,3, †], Cheng Tan[4], Yuanjun Yang[1], Zhifeng Zhu[1], Meixia Chen[1], Tingting Yan[1], Rong Hu[5], James Partridge[6], Guopeng Wang[5]*, Mingliang Tian[5], Ding-Fu Shao[2]* and Lan Wang[1,6]*

*1. Lab of Low Dimensional Magnetism and Spintronic Devices, School of Physics, Hefei University of Technology, Hefei, Anhui 230009, China;*

*2. Key Laboratory of Materials Physics, Institute of Solid State Physics, HFIPS, Chinese Academy of Sciences, Hefei 230031, China;*

*3 University of Science and Technology of China, Hefei 230026, China*

*4. Institute of Materials Research and Engineering (IMRE), Agency for Science Technology and Research (A*STAR), 2 Fusionopolis Way, Singapore 138634, Singapore*

*5. School of Physics and Optoelectronics Engineering, Anhui University, Hefei 230601, China;*

*6. School of Science, RMIT University, Melbourne, Victoria 3001, Australia*

[†]These authors contributed equally: Wei-Min Zhao, Yi-Lun Liu, Liu Yang

*e-mail: 20239@ahu.edu.cn; dfshao@issp.ac.cn; wanglan@hfut.edu.cn;

## Abstract

Magnetic tunnel junctions (MTJs) are crucial components in high-performance spintronic devices. Traditional MTJs rely on ferromagnetic (FM) materials but significant improvements in speed and packing density could be enabled by exploiting antiferromagnetic (AFM) compounds instead. Here, we report all-collinear AFM tunnel junctions (AFMTJs) fabricated with van der Waals A-type AFM metal $(Fe_{0.6}Co_{0.4})_5GeTe_2$ (FCGT) electrodes and nonmagnetic semiconducting $WSe_2$ tunnel barriers. The AFMTJ heterostructure device achieves a tunneling magnetoresistance (TMR) ratio of up to 75% in response to magnetic field switching. Our results demonstrate that the TMR exclusively emerges in the AFM state of FCGT, rather than during the AFM-to-FM transition. By engineering FCGT electrodes with either even- or odd-layer configurations, volatile or non-volatile TMR could be selected, consistent with an entirely interfacial effect. TMR in the even-layer devices arose by Néel vector switching. In the odd-layer devices, TMR stemmed from interfacial spin-flipping. Experimental and theoretical analyses reveal a new TMR mechanism associated with interface-driven spin-polarized transport, despite the spin-independent nature of bulk FCGT. Our work demonstrates that collinear AFMTJs can provide comparable performance to conventional MTJs and introduces a new paradigm for AFM spintronics, in which the spin-dependent properties of AFM interfaces are harnessed.

## Introduction

Magnetoresistive random access memory (MRAM) has emerged as a promising candidate for next generation information storage devices due to its unique combination of non-volatility, robust endurance, and potential for scalability[1-3]. The building block for MRAM is a magnetic tunnel junction (MTJ), typically composed of two ferromagnetic (FM) electrodes separated by a nonmagnetic (NM) barrier and utilizing tunneling magnetoresistance (TMR) to distinguish the relative magnetization orientations of the electrodes[4-8]. However, these MTJs suffer large stray fields and are further inhibited by the relatively slow dynamic responses inherent to ferromagnets. These factors hinder high-density and high-speed applications. Antiferromagnetic (AFM) materials, with vanishing stray fields and ultrafast dynamics have the potential to overcome these limitations[9-11]. However, the absence of a net magnetization implies a spin-independent nature of antiferromagnets, and hence an AFM tunnel junction (AFMTJ) has long been considered elusive[12]. Until very recently, AFMTJs have been constructed using two approaches, both of which rely on noncollinear AFM configurations. Electrodes that exhibit noncollinear AFM configurations and host the momentum-dependent transport polarization supportive to TMR are experimentally constructed in the first approach[13,14] (Fig. 1a). The second one uses a two-dimensional (2D) Van der Waals (vdW) AFM semiconductor as the barrier for a spin-filter AFMTJ[15], where a twisted stacking is introduced to generate a noncollinear alignment of the Néel vectors of the adjacent layer, allowing a magnetic field induced nonvolatile phase transition for a spin-filtering TMR (Fig. 1b). While these advances are significant and promising, an important gap remains in the field of AFMTJs: the realization of AFMTJs based on collinear AFM metals.

Collinear antiferromagnets, with two antiparallel-polarized sublattices, were discovered prior to their noncollinear counterparts[16]. However, a strong magnetoresistive effect in response to the rotation of Néel vector in collinear AFM devices has yet to be observed (Fig.1c). This stems from two major challenges: the lack of a collinear AFM system supportive to a convenient Néel vector manipulation, and the inability to generate spin-polarized transport in conventional antiferromagnets. A collinear AFMTJ with a large TMR would not only complete the spectrum of available AFMTJs but also assist in a deeper understanding of the fundamental physics underlying antiferromagnets. From a technological point of view, collinear AFMTJs, especially those utilizing a perpendicular magnetic anisotropy, possess significant advantages in scalability compared to noncollinear AFMTJs. This is due to the absence of magnetic canting at the edges of collinear AFMs.

## Results and discussion

Here, we present a purely collinear AFMTJ constructed as a Van der Waals (vdW)

heterostructure (Fig.1c). The recently discovered high-temperature vdW AFM metal $(Fe_{0.6}Co_{0.4})_5GeTe_2$ (FCGT) was selected for the device electrodes[17-21]. FCGT has a so-called A-type collinear AFM stacking, where the magnetic moments within each vdW layer exhibit FM alignment with perpendicular magnetic anisotropy, while the interlayer show AFM coupling, as illustrated in Fig. 2a. Since the A-type AFM materials with odd and even numbers of layers exhibit different properties in the 2D limit, we systematically investigated the transport properties of FCGT with varying numbers of layers (see Supplementary Material Fig.S1). As shown in Fig. 2c, a standard Hall device of even-layered FCGT (6L) exhibits near flat $R_{xy}$ between -1.8 T to +1.8 T, indicating its AFM magnetic ground state in this range. With an increase in the perpendicular magnetic field, the FCGT undergoes AFM-to-FM transitions around ±1.8 T and adopts a fully FM state at ±3.0 T, with a behavior generally found in vdW A-type AFM materials under an applied magnetic field. Hence, as shown in Fig. 2c, the FCGT states and the corresponding magnetic fields can be divided into three states; an AFM state (the blue region, up to 1.8 T), an AFM-to-FM transition state (the grey region, between 1.8 T and ~3.0 T), and a FM state (the yellow region, beyond ~3.0 T). In contrast to the relatively flat $R_{xy}$ in the AFM state of even-layered FCGT, a noticeable zero-field anomalous Hall effect is observed from the Hall device fabricated with odd-layered FCGT (see Supplementary Material Fig.S1).

The typical structure of the AFMTJ device is illustrated in Fig.2b, where two FCGT electrodes are separated by a semiconductor $WSe_2$ tunneling layer. To systematically investigate the AFMTJ, three types of devices were fabricated. These types differed in the numbers of layers in the incorporated FCGT. Specifically, the investigation was performed with even-even type (with both FCGT electrodes consisting of even-numbered layers), odd-even type (with odd-numbered FCGT layer in one side and even-numbered FCGT layer in the other side), and odd-odd type (with both FCGT electrodes consisting of odd-numbered layers) devices. Firstly, we focus on measurements obtained from the even-even configuration, named device D1, which exhibited the intrinsic characteristics of AFMTJs due to the absence of the net magnetization in even-layered FCGT. Atomic force microscopy (Supplementary Material Fig.S2) revealed that the top FCGT layer, middle $WSe_2$ layer, and bottom FCGT layer in device D1 had thicknesses of 6.0 nm, 2.8 nm, and 5.9 nm, corresponding to 6 L, 4 L, and 6 L structures, respectively. Given the critical role of the interfaces in MTJs, the pick-up method was used to fabricate the devices (see methods). In contrast to the metallic behavior observed from the FCGT standard hall device, the resistance-temperature curve of the AFMTJ devices shows insulating behavior (Supplementary Material Fig.S3), confirming that the $WSe_2$ layer was pinhole-free and functioned effectively as a tunneling barrier. We next focus on the characteristics of the AFMTJs.

Upon sweeping an out-of-plane magnetic field $\boldsymbol{B}$ within the range ±4T at 10K, a prominent TMR signal was observed in the AFMTJ with even-even configuration. As shown in Fig.2d, two prominent peaks emerged at ±0.9 T, which were within the AFM state and much lower than the AFM-to-FM transition field of FCGT. The highest resistance, $R_H$, of 61.2 kΩ was measured at +0.9 T and the lowest resistance, $R_L$, of 35.0 kΩ was measured at +1.5 T. Hence, the TMR ratio $(R_H - R_L)/ R_L$ reached 75 % at 10 K, comparable to that of FM-based vdW MTJs[22-27]. Such a large TMR ratio is unexpected in an AFMTJ due to the spin-degeneracy of an AFM. To eliminate the possibility that the observed TMR phenomenon arose from an AFM-to-FM phase transition, the emergence fields of the TMR and the AFM-to-FM transition fields of FCGT were collected from various devices and systematically compared (Supplementary Materials Fig.S5). The results show a significant gap between the two field regions, providing clear evidence that the TMR effect manifests in the AFM state. In addition to the prominent TMR behavior, a much smaller magnetoresistance transition occurred under a higher magnetic field of up to ±3.0 T (Supplementary Material Fig.S3b). This behavior is consistent with a change in the longitudinal magnetoresistance, $R_{xx}$, associated with the AFM-to-FM phase transition of FCGT. We note that this AFM-FM transition field is the same as that observed in the intrinsic FCGT hall device and consistent with those previously reported in FCGT[17]. This indicates that the AFM-to-FM transition behavior of FCGT is preserved in the FCGT/$WSe_2$/FCGT structure, supporting the assertion that the observed TMR behavior at +0.9 T originates from the AFM state of FCGT. The emergence of the prominent TMR signal within the AFM state and the negligible magnetoresistance change observed at the AFM-FM transition demonstrate that this TMR effect is an interface effect rather than a bulk effect. During the AFM-FM transition, although the net magnetic moment undergoes significant changes, the interfacial magnetic moment remains stable, leading to only minor fluctuations in magnetoresistance.

To systematically evaluate the performance of vdW AFMTJs, we performed temperature-dependent measurements on device D1. As shown in Fig. 3a, both the TMR ratio and the TMR transition field decrease with increasing temperature. To quantitatively display this phenomenon, we define TMR_max as the maximum TMR ratio and TMR_ZF as the TMR ratio at zero field. As shown in Fig.3b, a typical volatile TMR behavior is observed, which decreases monotonically with increasing temperature and vanishes above 230K. To further evaluate the interlayer interaction, we define $B_{C1}$ and $B_{C2}$ as the critical transition fields where the TMR effect respectively emerges and then vanishes under increasing magnetic field. As illustrated in Fig. 3c, at 10 K, the transition fields $B_{C1}$ and $B_{C2}$ exhibit values of 0.84T and 0.98T, respectively. Both fields show a gradual decrease with increasing temperature and ultimately vanish at high temperatures. Another even-even device, designated as D5 with 14L FCGT/ 5L $WSe_2$/ 8L FCGT, demonstrates analogous characteristics (see Supplementary Materials Fig.S4). We speculate that the interfacial moments serve to switch the Néel vector in FCGT. It should be noted that the interfacial magnetic moments only experience an interlay magnetic coupling of half intensity compared with the magnetic moment of inside layers. As illustrated in Fig.3d, when the magnetic field is increased to $B_{C1}$, the interfacial moments of one FCGT electrode are flipped first due to the relatively weak interlayer magnetic coupling, causing the Néel vector to switch due to interlayer AFM coupling. When the magnetic field is further increased to $B_{C2}$, this process occurs in the opposite FCGT electrode. Therefore, $B_{C1}$ and $B_{C2}$ can be considered as the "coercivities" of the interfacial moments. The Interface-assisted Néel vector switching is governed by the competition between magnetocrystalline anisotropy energy, AFM exchange energy, Zeeman energy, and domain wall dynamics. Although the two FCGT electrodes in device D1 have identical nominal thicknesses, differences in their local environments—arising from proximity coupling to distinct substrate or capping layers—and variations in defect distributions can lead to different effective anisotropies and domain pinning strengths, resulting in distinct critical switching fields. This asymmetry enables the observation of TMR. We note that such interface-assisted Néel vector switching has also been reported in other A-type antiferromagnets such as $Cr_2O_3$[28]. In this scenario, the TMR characteristics of the AFMTJs with even- and odd- layered FCGT electrodes should then differ, since their switching behaviors are different.

To explore this assertion, an odd-even type FCGT/$WSe_2$/FCGT device, named D2, was fabricated. Optical and atomic force microscopy (Supplementary Material Fig.S2) showed that the underlying FCGT layer number was 14 L and the upper FCGT layer number was 5 L. Hence, zero-net magnetic moment existed in one side and a residual net magnetic moment existed in the other side. As shown in Figs.4a and 4b, typical volatile TMR behavior was observed in this odd-even configuration below 80 K. Interestingly, as the temperature increased, non-volatile TMR behavior appeared above 80 K with TMR_max coincident with TMR_ZF in Fig. 4b, despite the absence of a pinning layer. The critical transition fields $B_{C1}$ and $B_{C2}$ were extracted to deduce the mechanism. Figure 4c shows $B_{C2}$ gradually approaching zero with increasing temperature, as that observed in the even-even AFMTJ devices and indicating that $B_{C2}$ is related to the interfacial coercivity of the even-layered FCGT electrode in this device. On the other hand, $B_{C1}$ is associated with the interfacial coercivity of the odd-layered FCGT. As the temperature increases, $B_{C1}$ first decreases and then clearly increases toward the opposite direction, with a sign change at 80 K. This behavior, unprecedented in conventional MTJs, is responsible for the non-volatile TMR at high temperatures.

The temperature-dependent non-volatile TMR behavior and $BC_1$ sign change behavior suggest two competing interactions[29-33]. Previous report suggests that during ascending or descending magnetic fields from FM states, $(Fe_{0.56}Co_{0.44})_5GeTe_2$ exhibits an intermediate state

M1. As shown in Fig. 4d and Fig. S7, the magnetic moments of all adjacent layers are antiparallel in the AFM alignment while the magnetic moments of the interface and its adjacent layers are aligned parallel in the M1 state[18]. In the process of descending field from positive to negative, as shown in Fig. S7b, the magnetic moments in all layers point to positive direction (the FM state) at large positive field. With decreasing field, FCGT first enter into the M1 state, where the energy of M1 state is lower than that of AFM state due to the lower Zeeman energy. Further decreasing the field, the energy of AFM state will become lower than that of M1 state. However, there is an energy barrier between the M1 state and the AFM state, either further decreasing the magnetic field or increasing thermal agitation energy are required to overcome the barrier and then drive FCGT from the M1 state to the AFM state. As shown in Fig. S7b, the descending field needs to get to negative to drive the M1 state to the AFM state at low temperature. At higher temperature, the AFM state can be realized at a positive field due to the larger thermal agitation energy. Hence, the interfacial coercivity $B_{C1}$ exhibits a negative value when $T < 80$ K. At higher temperature, the thermal energy is large, leading to a positive $B_{C1}$. The scenario shown here can also be viewed as an exchange bias between uncompensated interfacial FM monolayer and compensated AFM slabs, which can be further verified in the isolated odd-layered FCGT hall device (see Figure S9b). The complete spin switching diagram for the odd-even type device is depicted in Figure 4d, where the Neél vector switching occurs exclusively in the even layer, while the spin of the odd layer switches between the M1 alignment and AFM alignment.

To verify this physical picture, +1.5 T field-cooling and ±1.5 T field sweeps were performed on both the odd-even and even-even devices (Figure S8). Although 1.5 T exceeds the coercivity of the interface layer, it remains below the field required for the FCGT transition from AFM to FM states, thereby precluding the formation of the M1 magnetic alignment during negative field sweeps. As shown in supplementary material Figs. S8 and S9, the TMR peaks in the even-even device are symmetric during the ascending and descending magnetic field sweeps, consistent with the results from ±4T large-field scans. In contrast, the height and width of the TMR peaks from the odd-even device were asymmetric between the ascending and descending sweeps, attributed to the presence of the M1 alignment and an exchange bias[33] interaction in the odd-layered FCGT that results in self-pinning at high temperature. These results are in line with the half-loop transition in the schematic diagram in Figures 3d and 4d, verifying the switching of the Néel vector in the even layer and the switching of the M1/AFM alignment in the odd layer. The inherent properties of 2D vdW materials thus enable the realization of nonvolatile TMR signals in this system with no requirement for an extra pinning layer. We also fabricated odd-odd type devices (Figure S10) and since the M1/AFM transition occurs in both sides, only volatile TMR was observed.

The observation of conventional TMR behavior in even-even FCGT/$WSe_2$/FCGT and non-volatile TMR behavior in odd-even FCGT/$WSe_2$/FCGT is tantalizing, as it hints at a new underlying mechanism for TMR. Previous studies of TMR in AFMTJs have focused on the mechanisms associated with momentum-dependent[34,35] or sublattice-dependent[36,37] transport spin polarizations[12]. However, these two mechanisms can be excluded from our investigation of the behavior in the A-type collinear AFM FCGT, firstly because the FCGT slab maintains a preserved $PT$ symmetry—a combination of inversion symmetry ($P$) and time reversal symmetry ($T$)—leading to a spin-degenerate electronic structure. Secondly, the A-type stacking within the vdW structure suggests a negligible intra-sublattice coupling along the out-of-plane direction. Our experiments demonstrate that the interface plays a crucial role in supporting TMR and point to a distinct mechanism associated with the uncompensated interfaces[38].

To further elucidate this mechanism, first-principles quantum transport calculations were performed on vdW AFMTJs with A-type AFM electrodes. For illustrative purposes and without compromising generality, we adopted an AFMTJ structure comprising a central vacuum layer serving as the tunnel barrier, with artificial A-type AFM $Fe_5GeTe_2$ (A-F5GT) electrodes. This device structure, as depicted in Figures 5a and S11, allowed us to explore the physics underlying the TMR mechanism. The A-F5GT is designed to host fully occupied atomic sites and ordered Fe-vacancies based on $Fe_5GeTe_2$ (Figures 5a and S11), keeping $PT$ symmetry and A-type AFM properties similar to those of FCGT (Fig. 2a). This circumvents the computational challenge of simulating disordered split sites and doping atoms in FCGT. The $PT$ symmetry in A-F5GT enforces a spin-degenerate electronic structure (Fig. 5b). As a result, the number of conduction channels for the out-of-plane direction is identical for up-spin ($\sigma = \uparrow$) and down-spin ($\sigma = \downarrow$) electrons at any transverse wave vector $k_{\parallel} = \left(k_x, k_y\right)$ (Supplementary Material Fig. S11), implying the absence of bulk-driven spin polarized transport in this A-type antiferromagnet.

We define the P and AP states in such an AFMTJ according to the relative orientations of the uncompensated interfacial moments, and calculate the associated $k_{\parallel}$-dependent transmissions $T_P^{\sigma}(\vec{k}_{\parallel})$ and $T_{AP}^{\sigma}(\vec{k}_{\parallel})$ (Fig. 5c). We find $T_P^{\uparrow}\left(\vec{k}_{\parallel}\right)$ and $T_P^{\downarrow}\left(\vec{k}_{\parallel}\right)$ differ significantly, implying a finite transport spin polarization in the P states (Fig. 5c). On the other hand, we find $T_{AP}^{\uparrow}\left(\vec{k}_{\parallel}\right)$ and $T_{AP}^{\downarrow}\left(\vec{k}_{\parallel}\right)$ are significantly suppressed and become similar. This behavior resembles that of conventional MTJs with identical FM electrodes, and cannot be attributed to the bulk properties of the AFM electrodes. Rather, it stems from interface-driven spin-polarized tunneling. As a result, the total transmission for the P state ($T_P$) is much larger than that for the AP state ($T_{AP}$), leading to a TMR ratio $TMR = \frac{T_P - T_{AP}}{T_{AP}}$ as high as ~1000%. We find that replacing the vacuum layer with a realistic 2D insulator does not qualitatively influence the tunneling behavior of such an AFMTJ (Supplementary Material Fig. S12). The TMR ratio predicted by our calculation is larger than that obtained from our experiments. This discrepancy may be attributed to the significant disorder arising from the randomly distributed split sites and dopants in the selected FCGT samples. These introduce additional scattering effects not accounted for in our calculations, and hence significantly suppress the TMR ratio, especially at high temperature. With this in mind, enhanced TMR and higher operational temperatures may be made possible in AFMTJs by utilizing vdW A-type AFM electrodes with more ordered structures.

We note a recent study reporting TMR in MTJs with $(Fe_{0.8}Co_{0.2})_3GaTe_2$ (FCGaT) electrodes and $WSe_2$ barriers[39]. Crucially, this TMR originated from a field-induced AFM-FM phase transition, where high-resistance states emerge when one electrode remains AFM while the other becomes FM. This is in contrast to our system in which the FCGT exhibits TMR in the all-AFM state without bulk FM phase involvement and with inherent elimination of stray fields – a critical advantage for high-density integration. The divergent behaviors between FCGT and FCGaT likely stem from differences in exchange coupling strengths, magnetic anisotropy energies, interfacial cleanliness, and transfer methodologies—aspects warranting future investigation.

Our work demonstrates that uncompensated interfaces fulfill dual critical functions. They generate spin-polarized tunneling currents for magnetoresistive read-out and enable direct bulk Néel vector control for efficient write-in operations in all-AFM spintronic devices. A promising avenue for future exploration would be to use ultrafast optical method to directly detect the Néel vectors switching to study AFM dynamics and to manipulate the Néel vectors with the interface-assisted spin-orbit torque for all-electric AFM spintronic devices, as suggested recently[28,40,41]. Since most antiferromagnets exhibit uncompensated interfaces along selected growth directions, our approach suggests the possibility to exploit antiferromagnets previously viewed as “spin-independent” for spintronics, establishing a new device engineering paradigm.

In summary, we demonstrate that all-collinear AFMTJs can be successfully constructed as vdW heterostructures using standard techniques. These devices have achieved a TMR comparable to that of conventional MTJs. The A-type AFM stacking of the vdW electrodes facilitates convenient switching and supports nonvolatile TMR with no need for an additional pinning layer. The performance of these AFMTJs is largely unaffected by the thicknesses of the electrodes and barriers. These findings directly corroborate the theoretical proposal of interface-controlled collinear AFMTJs[38] and distinguish them from existing noncollinear AFMTJs. From a technological point of view, these exceptional characteristics, combined with the inherent advantages of antiferromagnets, herald significant potential for next-generation MRAM applications.

## Methods

### Crystal:

Single crystal FCGT was grown by the chemical vapor transport method. Full details can be found in our earlier work[17]. Single crystal $WSe_2$ was purchased from HQ graphene.

### Device fabrication and transport measurements:

The FCGT and $WSe_2$ flakes were exfoliated onto $SiO_2$(300nm)/Si substrates using blue tape. Ti/Au (5nm/25nm) contacts were patterned by photolithography and e-beam lithography and deposited by magnetron sputtering. Finally, devices were fabricated based on dry transfer method[42]. The FCGT, $WSe_2$, and FCGT flakes were sequentially picked up by polydimethylsiloxane/polycarbonate (PDMS/PC) and then transferred onto $SiO_2$/Si chips supporting the pre-patterned Ti/Au electrodes. The entire exfoliation and transfer process was carried out under inert conditions in an Ar-filled glovebox with $O_2$ and $H_2O$ concentrations below 0.1 ppm.

Transport measurements were made in a Physical Property Measurement System (PPMS) from Quantum Design, using an SRS Model CS580 and SR830 lock-in amplifier with a measurement frequency of 13.333Hz.

### Data processing for Hall measurements:

Due to the non-symmetry of the Hall device, the Hall resistance $R_{xy}$ was mixed with a small part of the longitudinal resistance $R_{xx}$. We processed the $R_{xy}$ data by using the standard symmetrical formula $(R_{xy1}(+\mathbf{B}) - R_{xy2}(-\mathbf{B}))/2$ to eliminate the contribution from $R_{xx}$, where $R_{xy1}$ and $R_{xy2}$ represent the Hall resistances measured during descending and ascending magnetic field sweeps, respectively, and **B** denotes the applied magnetic field. The TMR data are raw data and have not undergone symmetry processing.

### First-principles calculations:

First-principles calculations of the electronic structure were performed within the density functional theory (DFT)[43] using the projector augmented-wave (PAW)[44] method implemented in the VASP code[45,46]. The plane-wave cut-off energy was 500 eV. The k-point meshes used in the calculations were 12 × 12 × 6 for bulk $Fe_5GeTe_2$ and 12 × 12 × 1 for the vdW heterostructures. The exchange and correlation effects are treated within the generalized gradient approximation (GGA) developed by Perdew-Burke-Ernzerhof (PBE)[47]. The semiempirical DFT-D3 method parameterizing the van der Waals correction was used in the calculations[48].

The non-equilibrium Green's function formalism (DFT+NEGF approach)[49,50] was used to calculate quantum transport, which is implemented in QuantumATK[51]. The cut-off energy of 75 Ry, the nonrelativistic Fritz-Haber Institute (FHI) pseudopotentials, and $\vec{k}$-point meshes of 11 × 11 × 101 for $Fe_6GeTe_2$-based AFMTJs were used for self-consistent calculations. The $\vec{k}$-resolved transmission was calculated using 401 × 401 $k$-points in the two-dimensional (2D) Brillouin zone (BZ).

## Data availability

The data supporting the findings of this study are available from the corresponding author upon reasonable request.

**Acknowledgements**

L.W. was supported by the National Natural Science Foundation of China (Grant No. 12374177), L.Y. and D.F.S was supported by the National Key Research and Development Program of China (Grant Nos. 2024YFB3614101), the National Science Foundation of China (NSFC Grants No. 12274411, 12241405, and 52250418), the Basic Research Program of the Chinese Academy of Sciences Based on Major Scientific Infrastructures (No. JZHKYPT-2021-08), and the CAS Project for Young Scientists in Basic Research (No. YSBR-084). W.-M. Z was supported by the Anhui Provincial Natural Science Foundation (Grant No. 2408085QA011). The calculations were performed at Hefei Advanced Computing Center.

**Author contributions**

L.W. and D.-F.S. conceived the project. Y.-L.L. and W.-M. Z. fabricated the devices, performed the transport measurements and data analysis, assisted by C.T., Y.Y., Z.Z., M.C., T.Y., and L.W.. R.H., G.W. and M.T. synthesized the FCGT crystals. L.Y. and D.-F. S. performed the theoretical analyses and calculations. W.-M.Z., D.-F. S., J.P., and L.W. wrote the manuscript with assistance from all authors.

**Competing interests**

The authors declare no competing interests.

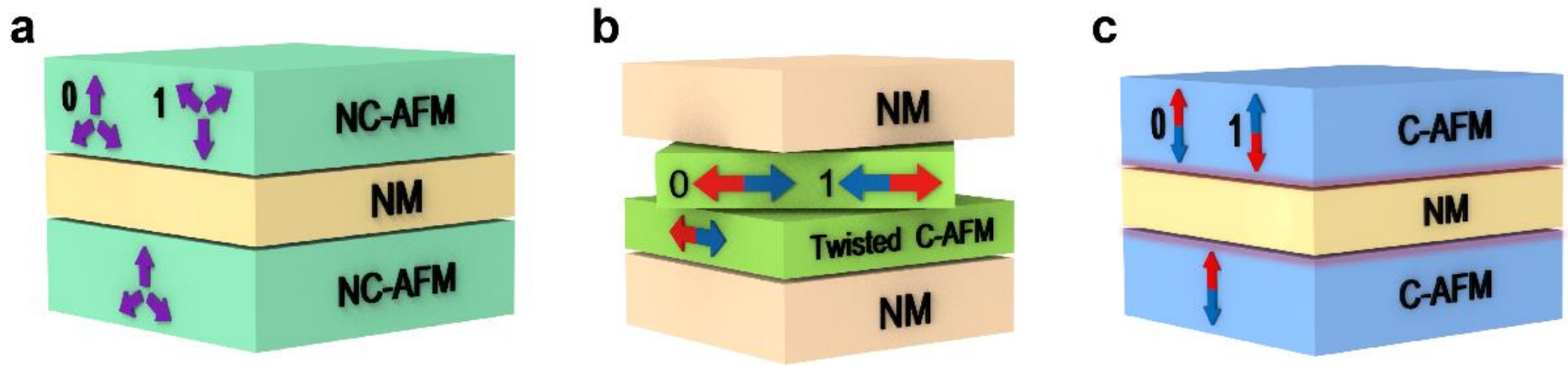

**Fig. 1 | Schematic of AFMTJs. a**, An AFMTJ based on non-collinear AFM (NC-AFM) electrodes and a non-magentic (NM) barrier. **b**, A spin-filter AFMTJ based on NM electrodes and an AFM barrier, where a non-collinear alignment of the Néel vectors is induced by twisting the collinear AFM (C-AFM) layers in the barrier. **c**, An all-collinear AFMTJ with C-AFM electrodes and an NM barrier. The interfaces are highlighted by red color.

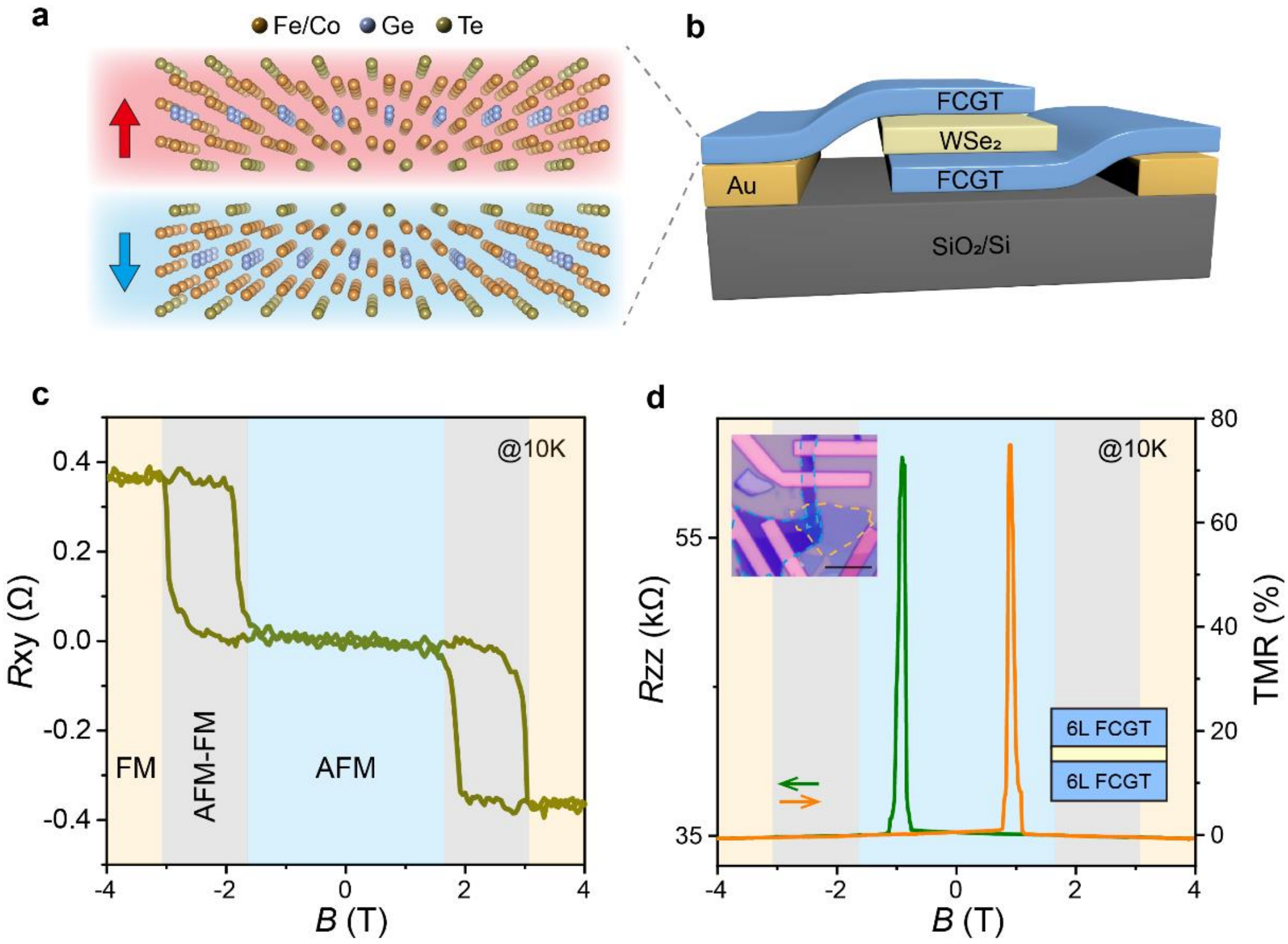

**Fig. 2 | TMR in the FCGT/$WSe_2$/FCGT AFMTJ. a**, Schematic of A-type AFM stacking of FCGT. **b**, Schematic of a FCGT/$WSe_2$/FCGT AFMTJ. **c**, Symmetrical Hall resistance $R_{xy}$ as a function of perpendicular magnetic field in 6 L FCGT at 10 K. The AFM state is colored by blue. The AFM to FM transition is colored by grey. The FM state is colored by yellow. **d**, Tunneling resistance $R_{zz}$ and corresponding TMR ratio for FCGT/$WSe_2$/FCGT AFMTJ device D1 at a fixed current 20 nA at 10 K. The sweep directions of the magnetic field are indicated by green and orange arrows. An optical image of device D1 is shown in the top-left inset, where the FCGT is outlined by the blue dashed line and the $WSe_2$ is outlined by the orange dashed line. The black scale bar is 10 μm. The thicknesses of the FCGT layers in device D1 are denoted in the bottom-right inset.

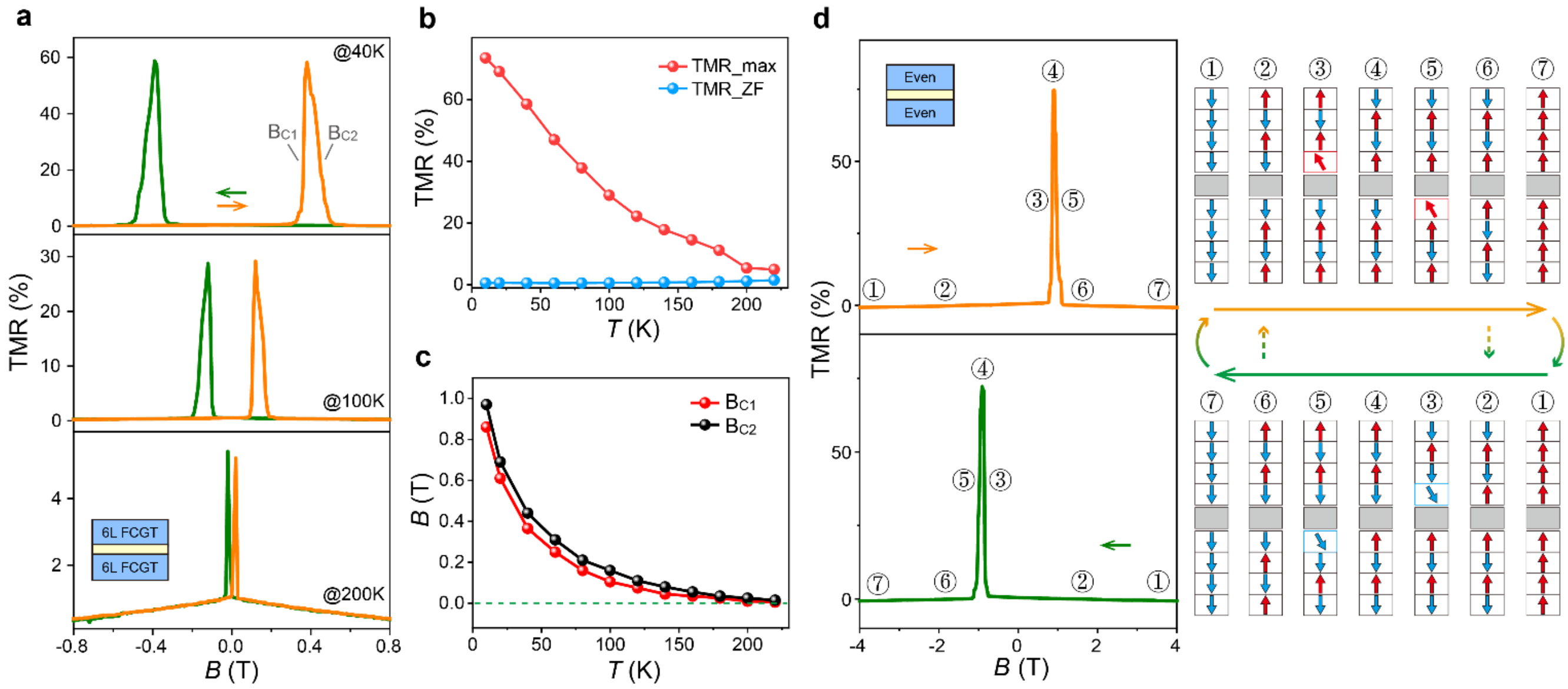

**Fig. 3 | The device performances and spin configuration switching diagram of the even-even AFMTJ. a**, Temperature-dependent TMR behavior of even-even device D1 at fixed current of 20 nA. The range of sweep field is ±4 T. The ascending and descending magnetic field sweeps are colored by orange and green, respectively. $B_{C1}$ and $B_{C2}$ are defined by the magnetic field at half peak height where the TMR effect respectively emerges and vanishes during field ascending. **b,** Temperature-dependent maximum TMR ratio (TMR_max) and TMR ratio at zero field (TMR_ZF). **c**, Temperature-dependent $B_{C1}$ and $B_{C2}$. **d**, Left: TMR ratio as a function of magnetic field for device D1 at 10K. Right: Schematic of spin configuration switching with applied magnetic field. A 4-layer spin configuration is used to represent even-numbered layers. The full-loop (±4T) scan is denoted by solid orange-green arrows. The half-loop (±1.5T) scan is represented by dashed orange-green arrows.

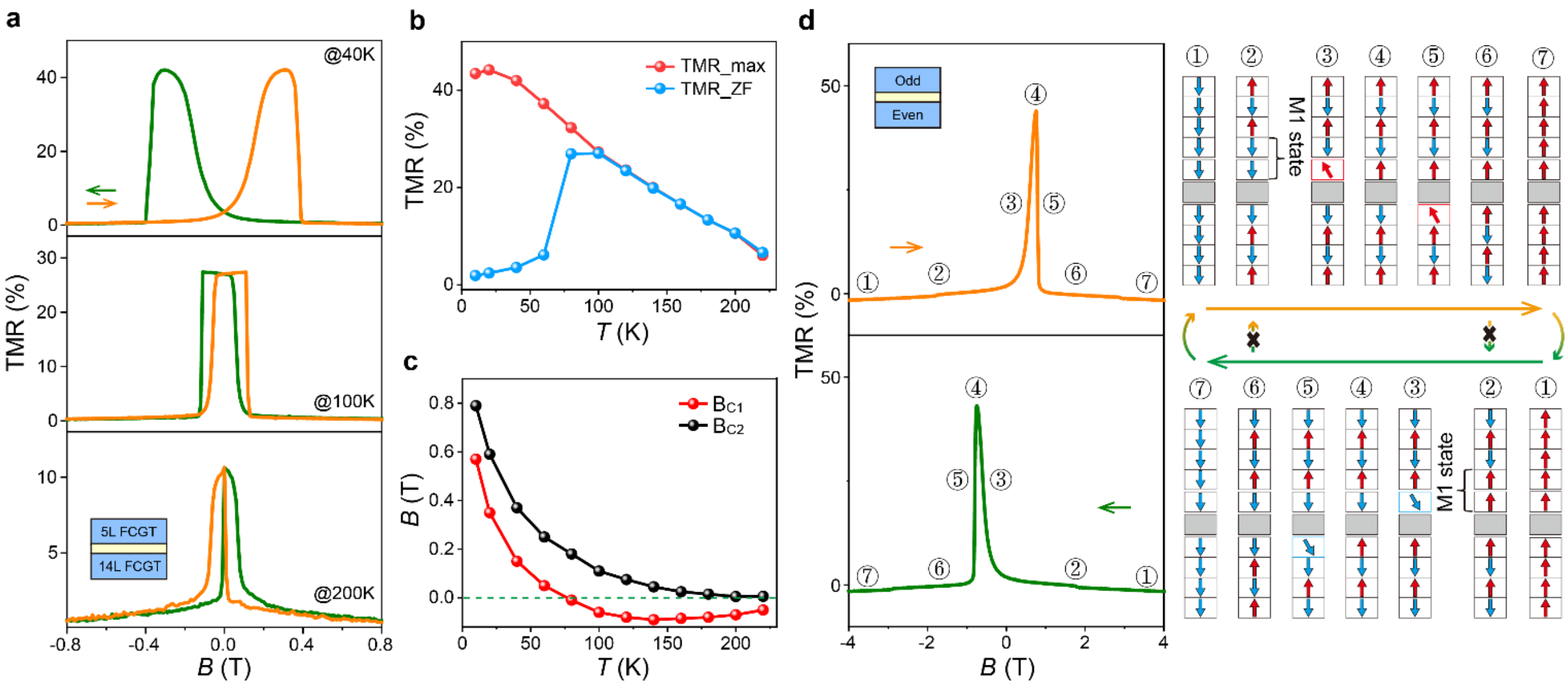

**Fig. 4 | The device performances and spin configuration switching diagram of the odd-even AFMTJ. a**, Temperature-dependent TMR behavior of odd-even device D2 at fixed current of 1 nA. The range of sweep field is ±4 T. The ascending and descending magnetic field sweeps are colored by orange and green, respectively. **b,** Temperature-dependent TMR_max and TMR_ZF. **c**, Temperature-dependent $B_{C1}$ and $B_{C2}$. **d**, Left: TMR ratio as a function of magnetic field for device D2 at 10K. Right: Schematic of spin configuration switching with applied magnetic field. Use a 5-layer spin configuration to represent odd-numbered layers and a 4-layer spin configuration to represent even-numbered layers. The full-loop (±4T) scan is denoted by solid orange-green arrows. The half-loop (±1.5T) scan is represented by dashed orange-green arrows.

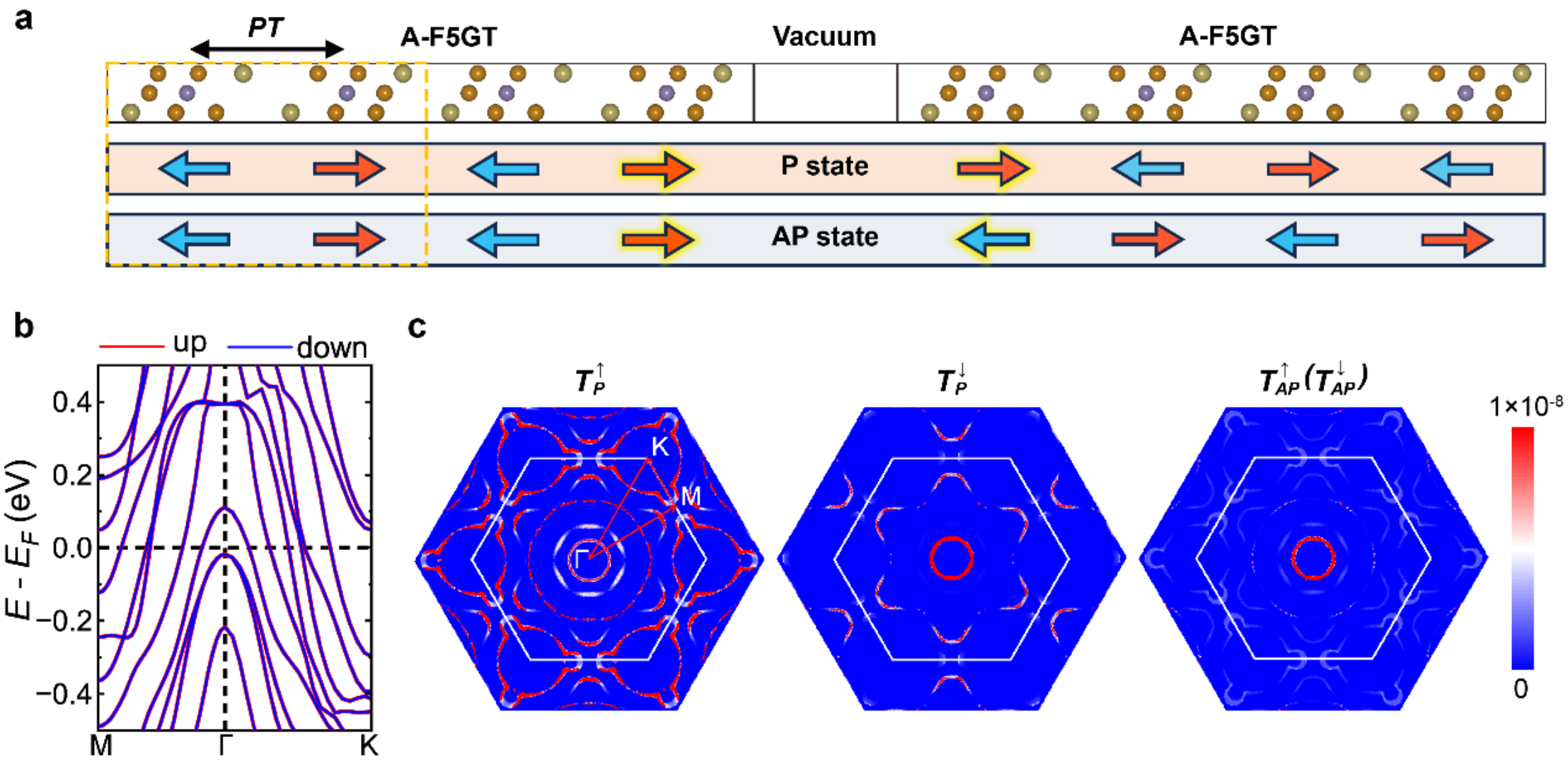

**Fig. 5 | Theoretical calculations of the interface-controlled vdW collinear AFMTJ.** **a,** Atomic structures of a A-F5GT/vacuum/A-F5GT AFMTJ where the left and right A-F5GT electrodes are considered to be of semi-infinite thickness in the ballistic transport model. The arrows indicate the magnetic moments alignments for P and AP states. The dashed box denotes the unit cell of A-F5GT electrode with a A-type AFM stacking. **b,** Band structure of bulk A-F5GT. **c,** Calculated $k_{\parallel}$-dependent transmissions $T_P^{\sigma}(\vec{k}_{\parallel})$ and $T_{AP}^{\sigma}(\vec{k}_{\parallel})$ at $E_F$ in the 2D Brillouin zone for P and AP states, where $\sigma$ denotes the spin channels.